\documentclass[twoside,twocolumn]{article}

\usepackage{blindtext} % Package to generate dummy text throughout this template 

\usepackage[sc]{mathpazo} % Use the Palatino font
\usepackage[T1]{fontenc} % Use 8-bit encoding that has 256 glyphs
\usepackage{microtype} % Slightly tweak font spacing for aesthetics

\usepackage[english]{babel} % Language hyphenation and typographical rules

\usepackage[hmarginratio=1:1,top=32mm,columnsep=20pt]{geometry} % Document margins
\usepackage[hang, small,labelfont=bf,up,textfont=it,up]{caption} % Custom captions under/above floats in tables or figures
\usepackage{booktabs} % Horizontal rules in tables

\usepackage{lettrine} % The lettrine is the first enlarged letter at the beginning of the text

\usepackage{enumitem} % Customized lists
\setlist[itemize]{noitemsep} % Make itemize lists more compact

\usepackage{abstract} % Allows abstract customization
\usepackage{titlesec} % Allows customization of titles
\renewcommand\thesection{\Roman{section}} % Roman numerals for the sections
\renewcommand\thesubsection{\roman{subsection}} % roman numerals for subsections
\titleformat{\section}[block]{\large\scshape\centering}{\thesection.}{1em}{} % Change the look of the section titles
\titleformat{\subsection}[block]{\large}{\thesubsection.}{1em}{} % Change the look of the section titles

\usepackage{fancyhdr} % Headers and footers
\usepackage{titling} % Customizing the title section

\usepackage{hyperref} % For hyperlinks in the PDF

\usepackage{color}
\usepackage{bbm}
\usepackage{tabularx, booktabs}
\usepackage[ruled,vlined]{algorithm2e}
\usepackage{array,multirow}
\usepackage{multirow,subcaption,caption}
\usepackage{afterpage}
\usepackage[dvipsnames]{xcolor}
\usepackage{enumitem}
\usepackage{setspace}
\usepackage{amsmath,graphicx}

\newcommand{\HarmonicMean}{\text{H}}
\newcommand{\Percentile}{\text{Perc}}
\newcommand{\Selection}{\text{PercSel}}
\newcommand{\queries}{\mathcal{Q}}

\newcommand{\Spearman}{\rho_s}
\newcommand{\Correlation}{\text{Corr}}
\newcommand{\Recall}{\text{Recall}}
    
\newcommand{\SelectPercentile}[3]{\ensuremath{\left\lceil#1\right\rceil_{#2, #3}}}
\newcommand{\population}{\ensuremath{\mathcal{X}}\xspace}

\newcommand{\feati}{\ensuremath{\boldsymbol{x}_i}\xspace}
\newcommand{\weights}{\ensuremath{\boldsymbol{w}}\xspace}
\newcommand{\pri}{\ensuremath{\boldsymbol{p}_i}\xspace}
\newcommand{\probout}{\ensuremath{\boldsymbol{c}}\xspace}
\newcommand{\PropGlobal}{\ensuremath{\boldsymbol\pi}}
\newcommand{\labels}{\ensuremath{\boldsymbol{y}}\xspace}
\newcommand{\likelihood}{\ensuremath{\boldsymbol{l}}\xspace}
\newcommand{\dth}{$\delta$\textsuperscript{th}\xspace}
\newcommand{\tth}{$\theta$\textsuperscript{th}\xspace}
\newcommand{\lth}{$\lambda$\textsuperscript{th}\xspace}
\newcommand{\ath}{$\alpha$\textsuperscript{th}\xspace}

\newcommand{\perceptron}{\hmm{s}tochastic gradient descent\xspace}
\newcommand{\naive}{\hmm{n}a\"{\i}ve\xspace}
\newcommand{\classifiersugg}{\ensuremath{\mathbb{C}_\text{1}}\xspace}
\newcommand{\classifierreg}{\ensuremath{\mathbb{C}_\text{2}}\xspace}

\newcommand\hmm[1]{\ifnum\ifhmode\spacefactor\else2000\fi>1000 \uppercase{#1}\else#1\fi}

\pretitle{\begin{center}\Huge\bfseries} % Article title formatting
\posttitle{\end{center}} % Article title closing formatting
\title{Inferring individual attributes from search engine queries and auxiliary information} % Article title
\author{%
\textsc{Luca Soldaini}\thanks{Work carried out during internship at Microsoft Research.} \\[1ex] % Your name
\normalsize Georgetown University, Washington D.C., USA\\ 
\normalsize \href{mailto:luca@ir.cs.georgetown.edu}{luca@ir.cs.georgetown.edu} 
\and 
\textsc{Elad Yom-Tov}\\[1ex] % Your name
\normalsize Microsoft Research, Herzeliya, Israel\\ 
\normalsize \href{mailto:eladyt@microsoft.com}{eladyt@microsoft.com} 
}
\date{} % Leave empty to omit a date
\renewcommand{\maketitlehookd}{%
\begin{abstract}
Internet data has surfaced as a primary source for investigation of different aspects of human behavior.
A crucial step in such studies is finding a suitable cohort (i.e., a set of users) that shares a common trait of interest to researchers.
However, direct identification of users sharing this trait is often impossible, as the data available to researchers is usually anonymized to preserve user privacy.
To facilitate research on specific topics of interest, especially in medicine, we introduce an algorithm for identifying a trait of interest in anonymous users.
We illustrate how a small set of labeled examples, together with statistical information about the entire population, can be aggregated to obtain labels on unseen examples.
We validate our approach using labeled data from the political domain.

We provide two applications of the proposed algorithm to the medical domain.
In the first, we demonstrate how to identify users whose search patterns indicate they might be suffering from certain types of cancer.
In the second, we detail an algorithm to predict the distribution of diseases given their incidence in a subset of the population at study, making it possible to predict disease spread from partial epidemiological data.
\end{abstract}
}

\begin{document}

% Print the title
\maketitle

%----------------------------------------------------------------------------------------
%	ARTICLE CONTENTS
%----------------------------------------------------------------------------------------

\section{Introduction}
\label{sec:introduction}

Identifying people with specific demographics, interests, or traits is a topic long of interest for researchers interested in online behavior and communities \cite{goel2012,hu2007}.
The ability to identify a cohort---that is, a group of people with a common defining characteristic---is a critical phase of the research process.
For example, when studying how the internet is used to seek medical advice, researchers have employed diverse heuristics to identify medical queries
\cite{palotti2016users,spink2004study} or health information seekers \cite{paul2016search,yomtov2014mood}.
Such heuristics are usually sufficient at identifying common queries and conditions, but fail to capture small cohorts, such as users suffering from an uncommon disease.
While such groups could be identified using personal information, demographic data, or health records, much of the data available to researchers is anonymous, in an effort to preserve the privacy of individuals.

In this paper, we introduce an algorithm for inferring individual attributes of a population of users by relying on a small set of examples with known labels and statistical information about the entire population. % (Section~\ref{sec:methodology}).
In other words, we show how to identify a cohort of interest by learning from a small set of users---which we identified using a very effective yet low-recall heuristic---and information about the distribution of the cohort of users in the entire population.

We validate the proposed algorithm by identifying the political affiliation of Twitter users: given a set of users and their tweets, we predict their political affiliation using a small set of users with known political orientation and statistics about the outcome of the elections. %  (Section~\ref{sec:evaluation}).
Our algorithm determines the affiliation of such users more effectively than other methods when the fraction of known users is small.

Finally, we present two applications of the proposed algorithm. % (Section~\ref{sec:applications}).
Both use the proposed system to create a cohort of users whose search patterns indicate they might be suffering from specific forms of cancer.
No personal data or patient history is used; rather, we combine a low recall, high precision heuristic with epidemiological data about incidence of the cancer.
Once the cohort has been determined, we show how to use it to train and evaluate a classifier to pre-screen for users who suffer from the cancer at study. % (Section~\ref{sub:applications:pre-screening}).
This shows how search queries can be used as a screening device for diseases that are often discovered too late, because no early screening tests currently exists.
Furthermore, we present a classifier that uses the cohort identified by the algorithm to predict the incidence of disease in regions where it is not known. % (Section~\ref{sub:applications:predicting-statistics}).
Such application could be useful to estimate the spread of a disease in regions where the number of reported cases is not sufficient to carry out a statistical analysis.

In summary, our contribution is threefold:
\begin{itemize}[noitemsep,nolistsep,topsep=0pt,leftmargin=*]
   \item We study the problem of \textbf{identifying cohorts of users who share a common trait} (e.g., they suffer from the same medical condition) from a population;
   \item  We \textbf{proposed and evaluate an algorithm} that couples fine-grained data on users with coarse-grained population statistics to identify cohorts for research purposes;
   \item We \textbf{describe and solve two possible applications} of the proposed algorithm: identification of users who might suffer from certain types of rare cancers and predicting the distribution of a disease in regions where it is unknown.
\end{itemize}

%!TEX root = resolving-missing-labels.tex

\section{Related Works}
\label{sec:related-works}

% TOPIC 0: background on medical applications

Traditionally, most of the medical research exploiting internet data  has focused on population-level disease incidence. The question therein are of the form ``\textit{how many people in a given area are currently suffering from influenza?}'' \cite{ginsberg2009}. Because of the large number of people involved,  it is superfluous to identify each individual with the condition. Instead, it is sufficient to find correlations between disease incidence and specific keywords \cite{paul2014,polgreen2008} or even website visits \cite{mciver2014}.

More recently, researchers have begun attempting to identify anonymous search engine users suffering from conditions of interest, either to provide individual level predictions or to learn from individual behaviors.
For example, Yom-Tov et al. \cite{yomtov2014mood} identified people suffering from mood disorders according to their queries of drugs used to treat the disorder, as well as changes in their behavior near the time of mood disorder events. In other work, Ofran et al. \cite{ofran2012patterns} used a threshold on the number of cancer-specific queries to identify people who were likely diagnosed with cancer and then track their information needs over time.
Good correlation was found between the number of people searching for cancer and disease incidence (but not prevalence) in the USA.
A more fine-grained approach was taken in Yom-Tov et al. \cite{yomtov2015} where a small subset of users was found to have identified themselves as suffering from a condition of interest.
The queries of this population were used to construct a classifier that predicted whether the condition a user was asking about most often was one they were suffering from.
The ability to identify users with specific conditions was then used to analyze their search histories for precursors of disease. More recently, Paparrizos et al. \cite{paparrizos2016} used people who self-identified as suffering from pancreatic cancer to predict their diagnosis ahead of time.

%Furthermore, researchers have investigated how the behavior of users changes as a response to a recent health event.
%Paul, et al. \cite{paul2015diagnoses,paul2016search} studied how the information need of searchers changes after being diagnosed with prostate cancer.
%De Choudhury, et al. \cite{dechoudhury2014characterizing} introduced a series of statistical models to estimate the likelihood of new mothers of being affected by post-partum depression by exploiting data collected before childbirth.

% TOPIC 1: ecological inference
The task of determining labels for individuals from population statistics relates to the ecological inference problem.
Ecological inference aims at inferring characteristics about individuals from ecological data (i.e., of the entire population).
As an example, it might be used to answer the following question: ``\textit{Given the number of votes for political parties A and B in a precinct and the number of men and women in the precinct, how many women voted for party A?}''
Ecological inference has a long history in the fields of statistics and social studies \cite{king1997solution}.
Recently, Flaxman, et al. \cite{flaxman2015who} used kernel embeddings of distributions to predict which demographics groups supported Barack Obama in the 2012 US Presidential Election.
Park and Gosh \cite{park2014ludia} introduced LUDIA, a low-level rank approximation algorithm designed that leverages ecological inference to predict hospital spending for individuals based on their length of stay.
Culotta, et al. \cite{culotta2016predicting} used website traffic data to predict demographics of Twitter user.
Ultimately, our problem differs from ecological inference in that we are interested in identifying individuals whose distribution is known rather than inferring behaviors at an individual level from population data.
% For example, in comparison to \cite{flaxman2015who}, we show how to determine the political affiliation of known users rather inferring voting behavior of demographics of the population.

% Culotta, et al. \cite{culotta2016predicting} proposed a method to predict aggregate demographic statistics

% TOPIC 2: learning with labeled proporsion
Another area of study that bears a similarity with our proposed algorithm is Learning with Label Proportions (LLP).
In LLP, the training data is provided to the classifier in groups on which only the distribution of classes in each group is known.
% The goal of the classifier is to learn a model to predict labels of individual instances.
Many solutions have been proposed for the problem  \cite{kuck2012learning,quadrianto2009estimating}; yet---to the best of our knowledge---none of them is designed to bias the learning process by incorporating individuals with known labels.
Keerthi, et al. \cite{selvaraj2011semi} introduced a semi-supervised SVM classifier that uses a small labeled dataset in conjunction to class proportion on the training data to predict labels on test data.
While sharing some similarity with our algorithm, their method is less generalizable, as it does not handle learning from training data drawn from sets with different class distributions.
Instead, our proposed approach solves this issue by conjunctively optimizing correlation with all sets the training data is drawn from.

Finally, many have studied semi-supervised learning (SSL), the problem of learning when a combination of labeled and unlabeled examples are available \cite{chapelle2006semi}.
For example, Druck, et al. \cite{druck2008learning} proposed a framework that leverages labeled features---that is, features that are highly representative for a class---to learn constrains for a multinomial logistic regression.
More recently, Ravi and Diao \cite{ravi2015large} have proposed a graph model to efficiently use SSL on large datasets.
Compared to a classic SSL model, we not only leverage individual level features, but also take advantage of population data.

% \begin{itemize}
% 	\item Geng, et al. \cite{geng2013facial} consider label distribution rather than a single label to predict, given a photograph, the age of the portrayed individual.
% 	\item Fr{\'e}nay and Verleysen \cite{frenay2014classification} present a survey for learning in the presence of class noise, i.e., noise on the labels.
% 	% \item Pietraszek \cite{pietraszek2007use} introduce an optimization technique for abstaining classifiers based on Receiving Operating Curve (ROC) analysis. In detail, three model are presented: the first takes advantage of explicit abstention cost, while the others implicitly model it.
% \end{itemize}

%!TEX root = resolving-missing-labels.tex

\section{Methodology}
\label{sec:methodology}

% 1. recap the goal of our system

% In this section, we detail the framework used to obtain labels on unseen examples.
% The framework combines information about a small set of examples whose labels are known with statistical information about the entire population from which the examples are drawn from.

% 2. introduce notation

\subsection{Notation}
\label{sub:methodology:notation}

% \begin{table}
%     \begin{tabularx}{\columnwidth}{|c|X|}
%         \hline
%         \textbf{Symbol}& \textbf{Meaning} \\
%         \hline
%         $\population$ & {Population of $n$ elements} \\
%         $X$ & {$n \times m$ features matrix} \\
%         $X$ & {Features Matrix} \\
%         $X$ & {Features Matrix} \\
%         $X$ & {Features Matrix} \\
%         \hline
%     \end{tabularx}
% \end{table}

Throughout this paper, we will adhere to the following notation:
scalars are identified by lowercase italic letters (e.g., $s$), vectors by lowercase bold letters (e.g., $\boldsymbol{v}$), and matrices by uppercase italic letters (e.g., $M$).
Calligraphy uppercase letters (e.g., $\mathcal{X}$) are used to denote sets.

Let $\population$ be a population of size $n = |\population|$.
To each element of $\population$, we associate the following: a features vector $\feati = \left\{x_{ij}\right\}_{j=1}^{m}$, a label $y_i \in \{0, 1\}$, and a property vector $\pri = \left\{p_{ik}\right\}_{k=1}^{t}$.
$y_i$ has value ``$1$'' if the $i$-th example belongs to the cohort of interest, ``$0$'' if its membership is unknown. We refer to the $n \times m$ matrix of all features vector as $X$.
A feature could be, for example, the use of a certain phrase by a user.
\pri represent a set of properties for an individual we directly take advantage in the proposed method.
For example, a property of an individual could be the US state where they are located;
in this case, \pri would be a $1 \times 51$ vector whose $k$-th position equals to ``$1$'' if the $i$-th individual is located in the $k$-th state, ``$0$'' otherwise.
While a property vector \pri is a feature vector for the $i$-th element of \population, it is convenient to consider it separately from $\feati$, as it simplifies the definition of the algorithm introduced in Section~\ref{sub:methodology:algorithm}.

We denote by \labels a $n \times 1$ vector holding all labels, while $P$ is a $n \times t$ matrix whose element $(i, k)$ represent the value of the $k$-th property for the $i$-th element of the population.

We encode the known statistical information as a $1 \times t$ vector $\PropGlobal$ containing statistical information about the property of individuals in $\mathcal{X}$.
For example, given a disease and a population of users located in the USA, $\PropGlobal$ could be a vector containing the incidence of the disease in each state.

Finally, we establish the notation for functions that will be used extensively in the reminder of the paper.
$\HarmonicMean(a, b)$ indicates the harmonic mean between the values $a$ and $b$.
We represent Spearman's rank correlation coefficient  between values of vectors $\boldsymbol{r}$ and $\boldsymbol{s}$ as $\Spearman(\boldsymbol{r}, \boldsymbol{s})$.
$\Percentile(\boldsymbol{r}, \alpha)$ returns the value in $\boldsymbol{r}$ corresponding to the \ath percentile;
building on the previous notation, we define the following operator:
\begin{equation}
    \Selection(\boldsymbol{r}, \alpha) = \{i\ |\ r_i \geq \Percentile(\boldsymbol{r}, \alpha)\quad \forall r_i \in \boldsymbol{r}\}
\end{equation}
$\Selection$ selects the set of indices of $\boldsymbol{r}$ whose corresponding values are in the \ath percentile.
The result of such function can be used to extract the matching components of any vector $\boldsymbol{s}$:
\begin{equation}
    \SelectPercentile{\boldsymbol{s}}{\boldsymbol{r}}{\alpha} = \{s_i\ |\ i\in \Selection(\boldsymbol{r}, \alpha)\}
\end{equation}
We will take advantage of the notation \SelectPercentile{\boldsymbol{s}}{\boldsymbol{r}}{\alpha} to identify the \ath percentile of vector $\boldsymbol{s}$ with respect to weight vector $\boldsymbol{r}$ throughout the manuscript.

\subsection{Proposed Algorithm}
\label{sub:methodology:algorithm}
\begin{algorithm}[t]
%\small
\linespread{1.0}\selectfont
\DontPrintSemicolon
\vspace{2pt}
 \KwData{Features matrix $X$, labels vector \labels, property matrix $P$, statistical information function $\PropGlobal$, number of iteration $\eta$, and learning percentile $\delta$.}
 \vspace{2pt}
 \KwResult{Vector $\likelihood = \{l_i\}$ of confidence values of element in $\population$ to be in the cohort of interest.}
\vspace{2pt}
\Begin{
    % $X \leftarrow \texttt{extractFeatures}(\population)$\;
    % $L \leftarrow \texttt{extractLabels}(\population)$\;
    % $P \leftarrow \texttt{extractProperties}(\population)$\;
    $\weights \leftarrow \texttt{initializeHyperplane}(\ )$; $o^{*} \leftarrow 0$\;
    \For{$j= 1, \ldots, \eta$}{
        $i \leftarrow \texttt{randomSample}(\{1, \ldots, n\})$; $\feati \leftarrow X[i]$\;
        $\boldsymbol{c}^{+} \leftarrow {(\weights + \feati)}/{||\weights + \feati||}$\;
        $\boldsymbol{c}^{-} \leftarrow {(\weights - \feati)}/{||\weights - \feati||}$\;
        $\boldsymbol{d}^{+} \leftarrow X \cdotp \boldsymbol{c}^{+}$; $\boldsymbol{d}^{-} \leftarrow X \cdotp \boldsymbol{c}^{-}$\;
        $o^{+} \leftarrow \HarmonicMean\left(\Correlation(\PropGlobal, P, \boldsymbol{d}^{+}, \delta), \Recall(\labels, \boldsymbol{d}^{+}, \delta)\right)$\;
        $o^{-} \leftarrow \HarmonicMean\left(\Correlation(\PropGlobal, P, \boldsymbol{d}^{-}, \delta), \Recall(\labels, \boldsymbol{d}^{-}, \delta)\right)$\;
        \If{$o^{+} > o^{-}$ \textbf{and} $o^{+} > o^{*}$}{
            $\weights \leftarrow \boldsymbol{c}^{+}$; $o^* \leftarrow o^{+}$\;
        }
        \ElseIf{$o^{-} > o^{*}$}{
            $\weights \leftarrow \boldsymbol{c}^{-}$; $o^* \leftarrow o^{-}$\;
        }
    }
    $\boldsymbol{d}^{*} \leftarrow X \cdotp \weights$\;
    $\likelihood \leftarrow \texttt{SoftmaxNormalize}(\boldsymbol{d}^{*})$\;
}
\linespread{1}\selectfont
\caption{The proposed SGD algorithm.}
\label{alg:perceptron}
\end{algorithm}

Recall that, given a population \population, we wish to identify a subset of \population---i.e., a cohort---such that all members of the cohort share a property of interest.
A solution for such problem should return a vector \likelihood of real values between 0 and 1 expressing the likelihood of each individual in \population of being part of the cohort of interest.
A \naive solution consists of using a classifier trained on the set of known members in the cohort.
However, as we will describe in Section~\ref{sub:evaluation-plan:results} and Table~\ref{table:performance-comparison}, this approach does not work well when the size of the set of users with known labels is small.

The algorithm we propose in this paper addresses this issue by conjunctively maximizing two quantities: (\textit{i}) the correlation between the counts of properties in the \dth percentile of users and the statistical information vector $\PropGlobal$ (e.g., the correlation between the number of users in the \dth percentile for each state and the incidence of the disease in each state), and (\textit{ii}) the fraction of known positive users (i.e., users whose label is ``$1$'') in the \dth percentile.
By optimizing for both quantities at the same time, we exploit the individual features of users that whose label is known, as well as statistical information about the distribution of the cohort of interest.

The algorithm works by finding a linear separating hyperplane which assigns a predicted label to each user given the features thereof.  
We formally define the two aforementioned quantities as follows: Let $\boldsymbol{d} = X \cdotp \weights$ be the vector of signed distances of elements from the decision hyperplane $\weights$;
then, \SelectPercentile{P}{\boldsymbol{d}}{\delta} is the $n' \times t$ property matrix associated with elements whose distances from \weights are in the \dth percentile.
In other words, \SelectPercentile{P}{\boldsymbol{d}}{\delta} contains the property vectors of those elements with distance from the decision hyperplane greater or equal than $\Percentile(\boldsymbol{d}, \delta)$.
Thus, we can define quantity ($i$) as:
\begin{equation}
    \Correlation(\PropGlobal, P, \boldsymbol{d}, \delta) := \Spearman\left(\PropGlobal,\ \mathbbm{1}^{1 \times n'} \cdotp \SelectPercentile{P}{\boldsymbol{d}}{\delta}\right)
\end{equation}
where $\mathbbm{1}^{1 \times n'}$ is the unit vector of size $1 \times n'$.

The fraction of known positive users (\textit{ii}) is the recall of the algorithm on known users; that is, the fraction of positive users whose distance from \weights is in the \dth percentile:
\begin{equation}
    \Recall(\labels, \boldsymbol{d}, \delta) := \frac{\left|\{y_i\ |\ y_i = 1\quad \forall y_i \in \SelectPercentile{\labels}{\boldsymbol{d}}{\delta}\}\right|}{\left|\{y_i\ |\ y_i = 1\quad \forall y_i \in \labels\}\right|}
\end{equation}
The two quantities determined by functions ``$\Correlation$'' and ``$\Recall$'' are combined by considering the harmonic mean of the two as the objective function.
We chose this mean as it penalizes the algorithm if the two quantities diverge significantly.

% The solution of the optimization problem detailed above can be achieved through an adaptation of the Perceptron classifier \cite{rosenblatt1958perceptron} that aims at finding a hyperplane \weights that separates elements from the positive and negative classes.

We used a modification of the Perceptron algorithm \cite{rosenblatt1958perceptron} in which a  stochastic gradient descent learns the hyperplane \weights separating elements of the positive and negative classes that maximizes the objective function described previously.

The details of the procedure are shown in Algorithm \ref{alg:perceptron}.
The algorithm iterates $\eta$ times over all elements in the population;
at each iteration, it randomly samples an element $i$;
then, it generates two candidate hyperplanes $\boldsymbol{c}^+$ and $\boldsymbol{c}^-$ by respectively adding and subtracting $\feati$ from \weights.
If any of the two candidate hyperplanes increases the value of the objective function, it then replaces \weights.
Finally, the confidence vector $\boldsymbol{d}^{*}$ is calculated by multiplying the feature matrix $X$ with \weights, and normalized by applying a Softmax normalization \cite{priddy2005artificial} to obtain the likelihood vector \likelihood.

%!TEX root = resolving-missing-labels.tex

\section{Validation}
\label{sec:evaluation}

% 1. we can't validate on medical data because we don't have enough medical data
% 2. however, we do have a dataset where users are divided in groups and on which we have statistical data
% 3. so we validate on those in this section

We validate the proposed algorithm on a dataset containing US Twitter users with known political affiliation.
This task has been studied in the past (e.g., \cite{barbera2015birds,conover2011predicting,culotta2016predicting}); in this paper, we use it as a benchmark for the proposed algorithm.

We show how the algorithm introduced in Section \ref{sub:methodology:algorithm}, when combined with statistical data on the outcome of the 2012 US presidential election, can be used to infer the political affiliation of users.
To do so, we hide a fraction $\gamma$ of users with known political affiliation by assigning them the label ``0'';
then, we measure the ability of the \perceptron in identifying these hidden users.

% The dataset was originally introduced in \cite{dyagilev2014linguistic}; we briefly describe it in Section \ref{sub:evaluation-plan:data-description}.
% Then, we report our experimental results in Section \ref{sub:evaluation-plan:results}.

\subsection{Data Description}
\label{sub:evaluation-plan:data-description}

Similarly to \cite{dyagilev2014linguistic}, we took advantage of a set of Twitter users with known political affiliation to evaluate our system.
Our dataset contains 372,769 users who explicitly expressed support for Barack Obama and 22,902 users who expressed preference for Mitt Romney during the 2012 US presidential election.
For the reminder of the paper, we will refer to the two groups as ``Democrats'' and ``Republicans'', while the set of all users will be identified as $\mathcal{U}$.

The political affiliation of members of $\mathcal{U}$ was determined by two sets of hashtags used by the supporters of the two parties during the election (e.g. ``\textit{\#romenyryan2012}'', ``\textit{\#voteobama}''; the complete list is available in \cite{dyagilev2014linguistic}).
This heuristic was found to have over 95\% accuracy \cite{dyagilev2014linguistic}.

The set $\mathcal{T}$ of all tweets generated by users in $\mathcal{U}$ between August 1\textsuperscript{st}, 2012 and November 15\textsuperscript{th}, 2012 was extracted.
We discarded all users for whom no location data was available (i.e., none of their tweets was geotagged), tweeted from two or more US states, or less than 30 times.
We identify the set of the $15,472$ remaining users ($900$ Republicans and $14,572$ Democrats) as $\population$.

We use the set $\mathcal{T}_{\population}$ of tweets associated with users in $\population$ to construct the feature matrix $X$.
For each user, we extracted the following features from their tweets: hashtags, mentions, domain name of URLs, and words occurring $10$ or more times in the corpus (except stopwords).
Prior works found such features to be effective at predicting the political affiliation of users \cite{barbera2015birds,conover2011predicting,culotta2016predicting,dyagilev2014linguistic}; in this work, we investigate their effectiveness when paired with the proposed algorithm.
% The hashtags used in \cite{dyagilev2014linguistic} to identify political affiliation of users were excluded from the feature set.
% Each feature is weighted by its frequency in all tweets by the user.

We represent the state each user in $\population$ belongs to through property matrix $P$;
in other words, $P$ is a $15,472 \times 51$ matrix where position $(i, k)$ is equal to $1$ if the $i$-th user tweets from the $k$-th state (according to geo-tagging), $0$ otherwise.

The population statistic vector $\PropGlobal$ was derived from the total count of votes casted for the Republican and Democrat candidates in each US state as disclosed by the official Federal Election Commission report \cite{fec2013elections}. %\footnote{\url{http://www.fec.gov/pubrec/fe2012/federalelections2012.shtml}}.
Specifically, the $k$-th value of $\PropGlobal$ represents the number of Republican voters for each inhabitant in the $k$-th state.
We normalized $\PropGlobal$ by the number of active users in each state within the time frame of data collection; this gave us the expected number of Republican Twitter users in each state.

\subsection{Results}
\label{sub:evaluation-plan:results}

\begin{figure}[t]
    \centering
    \includegraphics[width=0.75\columnwidth]{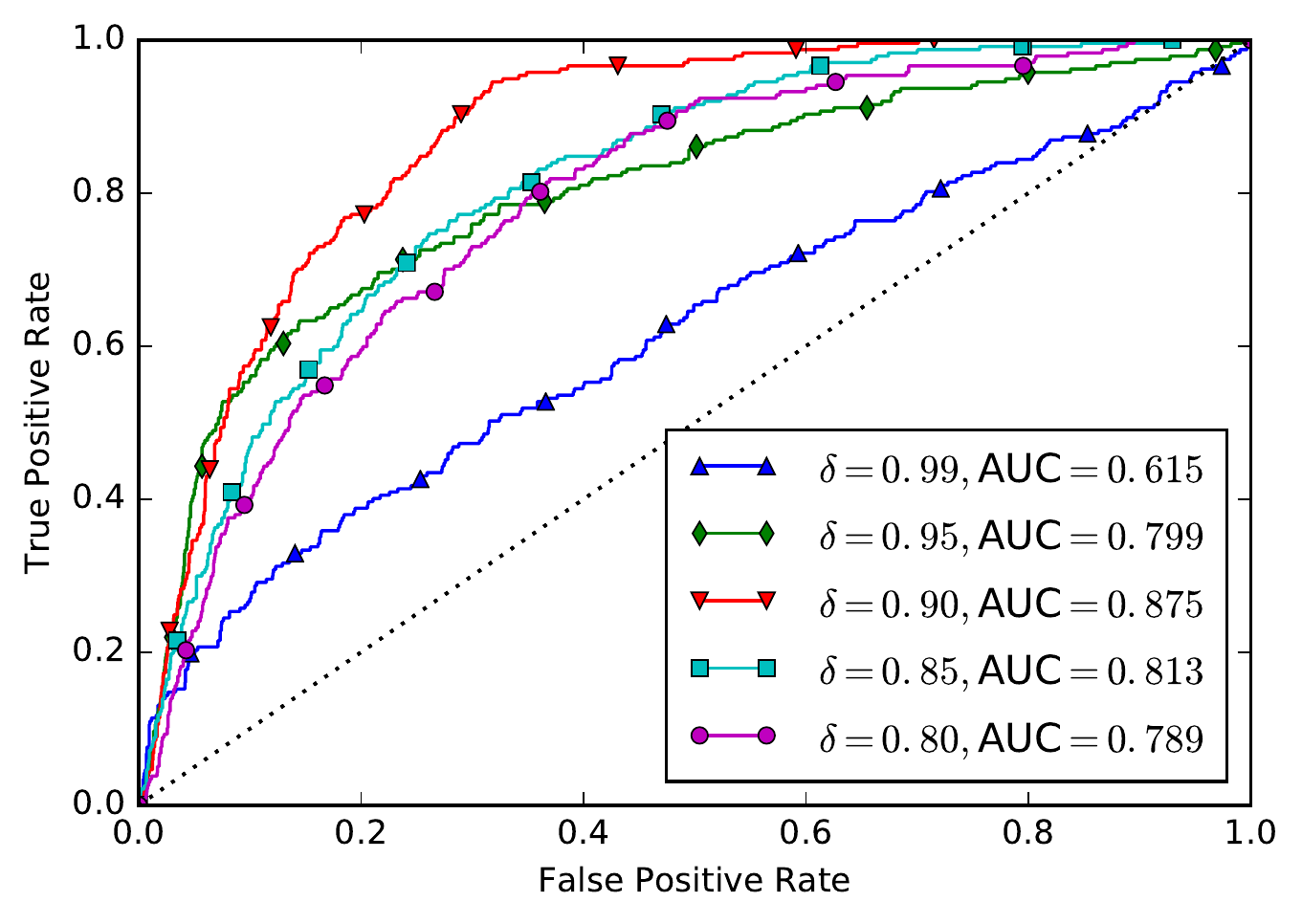}
    \caption{
    ROC curve of the \perceptron algorithm.
    The fraction of ``hidden'' Republicans is kept constant at ${\gamma = 0.75}$, while the learning percentile ${\delta}$ is varied.
    % The best AUC is obtained when ${\delta = 0.9}$;
    % in this case, the number of users classified as positive is close to fraction of Republicans in $\population$.
    }
    \label{fig:roc-delta}
    \vspace{-12pt}
\end{figure}

In this section, we illustrate the performance of the proposed algorithm in identifying Republican users whose label have been hidden.
Recall that, in order to quantify the ability of Algorithm \ref{alg:perceptron} to correctly label users in $\population$, we remove the label of a fraction $\gamma$ of republican users;
that is, we assign them the label ``0''.
Therefore, we tested the algorithm for various values of $\gamma$, as well as multiple values of learning percentile $\delta$.
The exploration of different parameters allows us to test how the algorithm behaves when number of users in the cohort is not known (if available, such number could be used to tune $\delta$).
All the experiments were executed under five-fold stratified cross validation; the number of iterations $\eta$ was set to $30,000$ to ensure reaching a stable point.

\begin{figure}[t]
    \centering
    \includegraphics[width=0.75\columnwidth]{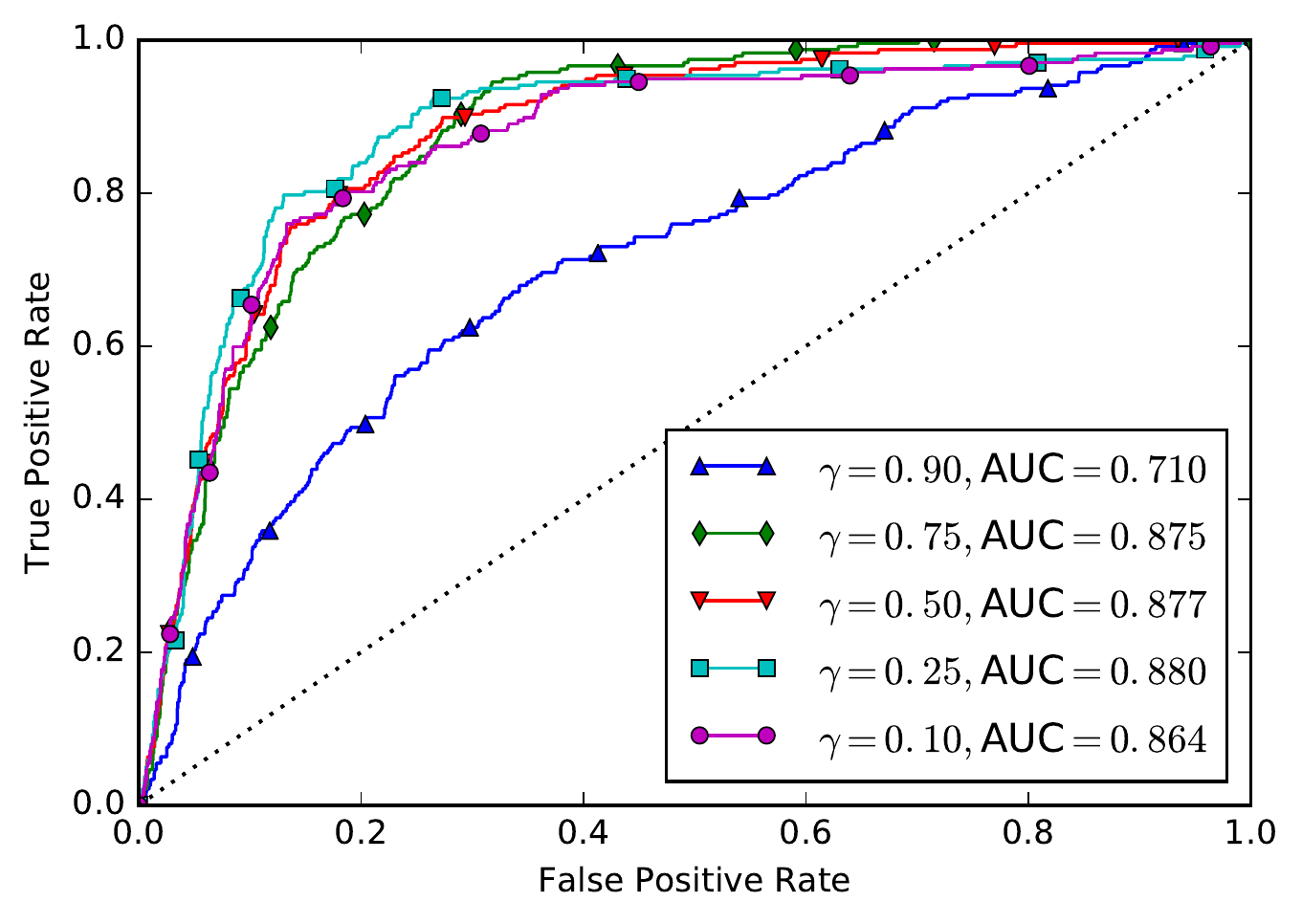}
    \caption{ROC curve of the \perceptron algorithm.
    The learning percentile is kept constant at ${\delta = 0.9}$, while the fraction of ``hidden'' Republicans ${\gamma}$ is varied.
    % The AUC increases as the number of disclosed Republicans increases;
    % however, minimal improvements are observed for $\gamma < 0.75$, meaning that good classification performances can be obtained with just a small set of known Republicans.
    }
    \label{fig:roc-gamma}
    \vspace{-12pt}
\end{figure}

We report the results of our experiments in Figures~\ref{fig:roc-delta} and \ref{fig:roc-gamma}, as well as in Tables~\ref{table:performance-comparison} and \ref{table:features-political}.
Specifically, Figure \ref{fig:roc-delta} shows the Receiver Operating Characteristic (ROC) curves produced by varying values of the learning percentile $\delta$.
For this experiment, $\gamma$ is fixed at $0.75$.

Two observations can be made about the results.
First, we note that the performance of the classifier, as measured by the Area Under the Curve (AUC), increases as $\delta$ approaches $0.9$;
then it starts declining.
We explain this behavior by observing that the number of elements in the $10$\textsuperscript{th} percentile is close to the number of Republican in $\population$;
therefore, as $\delta$ approaches this value, the performance of the proposed algorithm increase.
Past the $10$\textsuperscript{th} percentile, more noise is introduced, thus affecting the quality of classification outcome.
We remark that the decline in performance is not abrupt;
this characteristic is desirable, as in applications of the proposed algorithm (such as those shown in Section \ref{sec:applications}) the exact size of the cohort of interest is often unknown.

Second, we observe that classifiers with larger values of $\delta$ (e.g., $\delta = 0.95$) have a higher true positive rate associated with lower false positive rate (bottom left of Figure~\ref{fig:roc-delta}).
This is likely due to the fact that such classifiers make fewer mistakes on elements they have high confidence in (i.e., the values in \likelihood for with high-confidence elements are close to $1$).

For the second experiment (Figure \ref{fig:roc-gamma}) we varied the fraction of hidden users $\gamma$ between $0.9$ (i.e., only $10\%$ of Republicans are disclosed) and $0.1$ ($90\%$ of Republicans are disclosed).
We observe the AUC increases as $\gamma$ decreases;
that is to be expected, as less hidden republicans equals a more diverse pool of training examples.
However, to our initial surprise, we also noticed that the performance of the classifier show little improvement for values of $\gamma < 0.75$.
Such behavior is beneficial for the applications where this algorithm will be used, where typically only a small set of users of the cohort of interest is known.

We compared the proposed system with a simple Linear Stocastic Gradient Descent (LSGD), as well as with the Support Vector Machine (SVM) classifier proposed by Conover, et al. in \cite{conover2011predicting}.
For all three systems, we kept the ratio of hidden Republican users set to $\gamma=0.75$, as we were interested in studying how the proposed algorithm compares to other algorithms when only a small set of users in the cohort is known.
As shown in Table~\ref{table:performance-comparison}, the proposed algorithm outperforms both baselines, confirming that combining statistical information about the distribution of the cohort in the population with weights learned from individual features is an effective strategy to solve the task introduced in this paper.
For the two baselines, we experimented with using just the features in $X$ to train the classifiers, as well as concatenating $P \cdot \PropGlobal$ with the features matrix  $X$.
Interestingly, the performance of the two baselines decrease when augmenting $X$ with $P \cdot \PropGlobal$,  suggesting that the \naive approach of expanding the feature set with population statistics is not effective at identifying the cohort of users.

The features that were assigned the highest weights are the most  indicative phrases used by positive (Republican) users. We report features with the highest weight in $w$ for the classifier $\delta = 0.9, \gamma = 0.75$ in Table \ref{table:features-political}.
We note that the hashtags ranked in $1^{st}$, $3^{rd}$, $4^{th}$, $5^{th}$, $6^{th}$, $9^{th}$ places are typically used in right-wing circles;
the remaining hashtag (``\textit{\#landslide}'') while related to the election, is not unique to the rhetoric on any of the two political parties.
Finally, we note that the all URLs shown in Table \ref{table:features-political} are of websites leaning on the right side of the political spectrum.

\begin{table}[t]
    % \vspace{-6pt}
    \centering
    {\def\arraystretch{1.0}
	%\small
	\begin{tabular}{|c|c|}
        \hline
        \textbf{Classifier} & \textbf{AUC} \\
        \hline
        Linear Stochastic Gradient Descent (LSGD) & $0.667$ \\
        \hline
        LSGD + property vectors as features & $0.614$ \\
        \hline
        Support Vector Machine (SVM) from \cite{conover2011predicting} & $0.703$ \\
        \hline
        SVM \cite{conover2011predicting} + property vectors as features & $0.629$ \\
        \hline
        Proposed SGD ($\delta = 0.9$) & $\mathbf{0.875}$ \\
        \hline
    \end{tabular}
    \vspace{-8pt}}
    \caption{Comparison of the proposed algorithm to previously-proposed baselines. When a small fraction of Republican users is used for training ($\gamma = 0.75$), the algorithm outperforms a linear SGD baseline and the system from \cite{conover2011predicting} (difference is statistically significant, Wilcoxon signed-rank test, $p < 0.05$).}
    \label{table:performance-comparison}
    \vspace{-8pt}
\end{table}
\begin{table}[t]
    % \vspace{-6pt}
    \centering
    {\def\arraystretch{1.0}
	%\small
	\begin{tabular}{|c|c|c|}
        \hline
        \textbf{Rank} & \textbf{Feature} & \textbf{Weight} \\
        \hline
        1 & \#4moredays & $0.0517$ \\
        2 & \#landslide & $0.0490$ \\
        3 & \#loveofcountry & $0.0377$ \\
        4 & \#whyiamnotvotingforobama & $0.0244$ \\
        5 & \#whyimnotvotingforobama & $0.0229$ \\
        6 & \#bengahzi & $0.0148$ \\
        7 & anncoulter.com & $0.0129$ \\
        8 & searchnc.com & $0.0112$ \\
        9 & \#bengha & $0.0111$ \\
        10 & personalliberty.com & $0.0110$ \\
        \hline
    \end{tabular}
    \vspace{-8pt}
    }
    \caption{Top ten features for classifier ${\delta = 0.9, \gamma = 0.75}$.
    Websites ranked 7, 8, and 10 are right-leaning publications.}
    \label{table:features-political}
    \vspace{-8pt}
\end{table}

%!TEX root = resolving-missing-labels.tex

\section{Applications}
\label{sec:applications}

We present two applications of the algorithm introduced in the previous sections.
The first (Section \ref{sub:applications:pre-screening}) deals with identifying users of a search engine whose search patterns suggest a higher risk of developing a certain type of cancer; the second (Section \ref{sub:applications:predicting-statistics}) is concerned with predicting the incidence of two forms of cancer in regions of the USA.

We focus on ovarian cancer and cervical cancer. These relatively rare cancers (affecting approximately $12$ and $10$, respectively of $100,000$ women in the USA), are also quite deadly: Indeed, though ovarian cancer accounts for only 3\% of all cancers in women, it is the deadliest cancer of the female reproductive system \cite{us2013united}. One reason for this is that symptoms of these cancers are relatively benign, which means that many women are diagnosed at late stages of the cancer, though treatment is most effective in early stages. Additionally, no simple screening test is available for these cancers \cite{buys2011effect}. Thus, the ability to pre-screen for these cancers using Internet data could be of significant importance.

%For both applications, we use two datasets extracted from Bing query logs, which we will describe in Section \ref{sub:applications:data_description}.

%Finally, in Section \ref{sub:applications:similarity-self-identifed}, we quantify the difference between the subset of self-identified users in the population and users labeled by the proposed algorithm, confirming that self-identified users are not sufficiently representative of the population to be used by themselves.

\subsection{Data Description}
\label{sub:applications:data_description}
\begin{figure*}[th]
    \centering
    \begin{subfigure}[t]{0.78\textwidth}
        \includegraphics[width=1\textwidth]{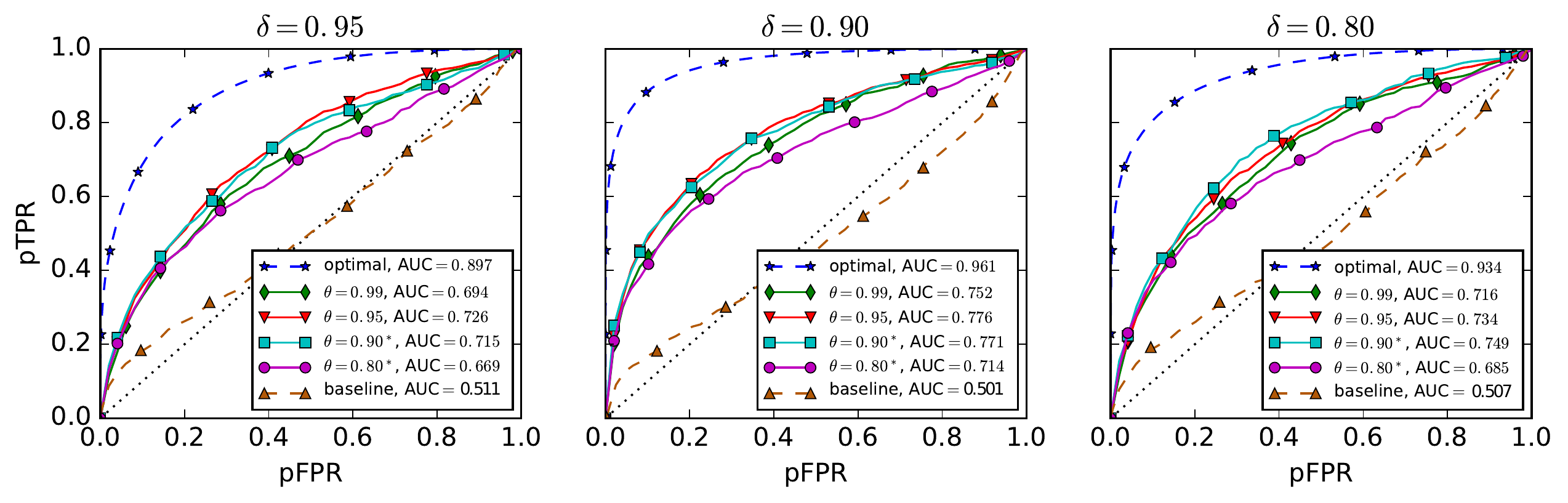}
        \caption{Ovarian cancer users}
        \label{subfig:ovarian-svm-flt}
    \end{subfigure}
    \begin{subfigure}[t]{0.78\textwidth}
        \includegraphics[width=\textwidth]{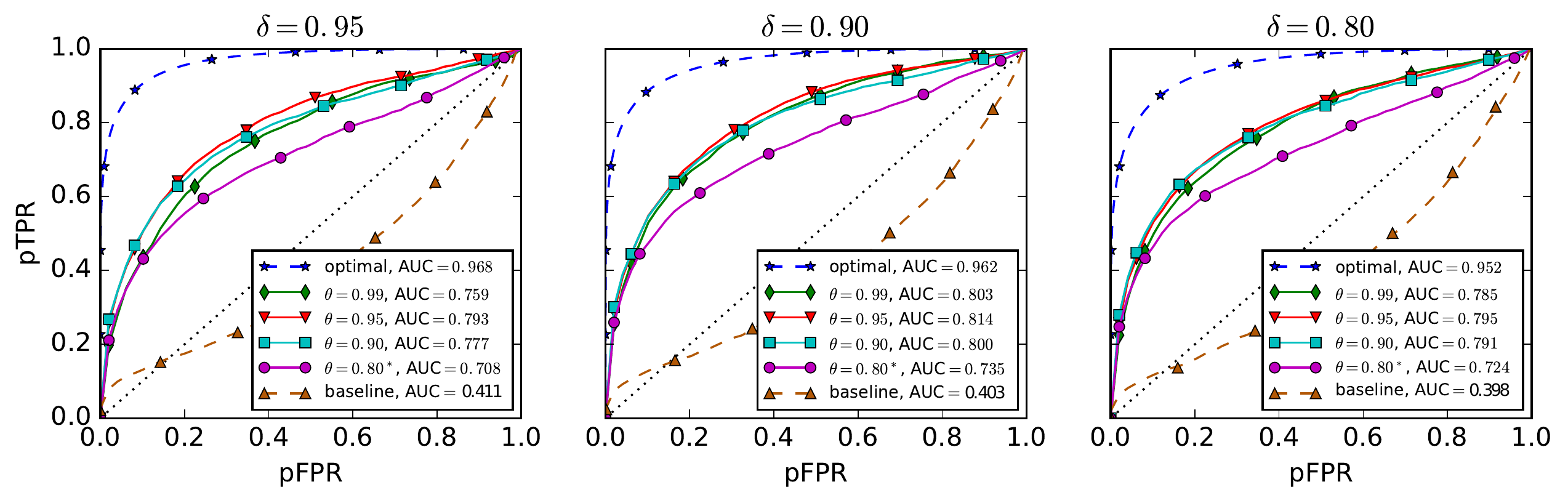}
        \caption{Cervical cancer users}
        \label{subfig:ovarial-svm-flt}
    \end{subfigure}
    \vspace{-4pt}
    \caption{
        pROC curves for ovarian cancer (Figure~\ref{fig:proc}a, top) and cervical cancer (Figure~\ref{fig:proc}b, bottom).
        The values of the learning percentile $\delta$ are reported above each figure.
        The AUC of \classifiersugg under multiple values of $\theta$ are shown alongside the optimal and baseline classifiers.
        Values of $\theta$ maked as * are statistically different from the best runs (Wilcoxon signed-rank test, $p < 0.05$).
    }
    \label{fig:proc}
    \vspace{-10pt}
\end{figure*}

\begin{table}[t]
    \centering
    %\small
    {\def\arraystretch{1.0}\begin{tabular}{|p{0.45\columnwidth}|p{0.45\columnwidth}|}
        \hline
        \multicolumn{1}{|c|}{\textbf{Features in }$\mathbf{\mathit{X}}$} & \multicolumn{1}{c|}{\textbf{Features in }$\mathbf{\mathit{Z}}$} \\
        \hline
        {list of ovarian/cervical \newline cancer symptoms$^{*}$} & {list of ovarian/cervical \newline cancer symptoms$^{*}$}\\
        \hline
        {$q$ most common terms \newline in queries $\queries$ \textbf{after} \newline disease mention}&{$q$ most common terms \newline in queries $\queries$ \textbf{before} \newline disease mention}\\
        \hline
        {list of symptoms$^{*}$} & {list of symptoms$^{*}$}\\
        \hline
        {list of diseases$^{*}$} & \\
        \cline{1-1}
        {list of names of drugs} & \\
        \cline{1-1}
        {list of names of US hospitals} & \\
        \hline
    \end{tabular}
    \vspace{-8pt}}
    \caption{Features used to construct matrices $X$ and $Z$.
    Features in $X$ are extracted from queries issued \textit{after} the first query mentioning the disease, while features in $Z$ are extracted from queries issued \textit{before} the first query mentioning the disease.
    $q$ was set to $2,000$ after empirical evaluation.}
    \label{table:features-applications}
    \vspace{-12pt}
\end{table}

Two sets of Bing users were considered to evaluate the applications introduced in this section. The first population, which we identify as $\population^{ov}$, consists of users who are likely to be suffering from  ovarian cancer.
%Our interest in studying this disease is due to the fact that, despite its low incidence (it only accounts for 3\% of all cancers in women), ovarian cancer is the deadliest cancer of the female reproductive system \cite{us2013united}.
%Furthermore, while treatment for ovarian cancer is the most effective in early stages, no screening test is available  yet \cite{buys2011effect}.
The second population of users is of users who are potentially suffering from cervical cancer.
%Such disease, while less deadly than ovarian cancer, has a similar incidence, thus being another useful dataset to validate the applications presented in this section.
We refer to this group as $\population^{cr}$.

We stress that it is traditionally very challenging to identify those users in $\population^{ov}$ or $\population^{cr}$ who are affected by the aforementioned conditions using Internet data, because both diseases have a low incidence rate, thus causing any heuristic---such as extracting all users who issue a specific query---to retrieve too few individuals.
Thus, we apply the algorithm introduced in Section~\ref{sec:methodology} to obtain an estimate of the probability of each user of suffering from cancer.

For the reminder of the paper, we will refer to the set of users $\population$ to describe all procedures that are common to both $\population^{ov}$ and $\population^{cr}$;
differences will be pointed out when necessary.

To obtain $\population$, we proceed as follows:
First, using the websites of the Center for Diseases Control (CDC) and the American Cancer Society, we produced a list of symptoms and drugs commonly associated with the each of the two diseases.
The two lists were expanded using two experts-to-laypeople synonym mappings, \textit{MedSyn} \cite{yates2013adrtrace} and \textit{Behavioral} \cite{yom2013postmarket}.
This expansion was made so as to bridge the gap between the vocabulary used by health experts and expressions preferred by laypeople \cite{soldaini2016enhancing}.
We extracted all Bing users in the United States who have queried in English in a span of five months: Ovarian cancer from April to August 2015 and cervical cancer from June 2015 to October 2015, for any of the symptoms or drugs associated with the cancer and the name of the cancer itself.
Finally, we extracted the US state of origin of each user through reverse IP address lookup to take advantage of the state-level incidence statistics for the two types of cancer.
Users who were associated with two or more US states were discarded.
This heuristic identified $3,167$ users who potentially have ovarian cancer and $9,327$ users who might have been diagnosed with cervical cancer.
Not all users in the two sets are affected by the respective diseases;
rather, the heuristic was used to reduce class imbalance before using the proposed algorithm to derive their likelihood of having the condition.
We refer to the set of all queries issued by all users in $\population$ as $\queries$.

For both conditions, we identify a set of users who are known (by their own admission) to be affected by cancer, as in \cite{yomtov2015}. This was done by finding all users who issued a query starting with ``\textit{i have }\texttt{<condition>}'' or ``\textit{i was diagnosed with} \texttt{<condition>}'', where \texttt{<condition>} is either ``\textit{ovarian cancer}'' or ``\textit{cervical cancer}''.
We will refer to these users as  ``self-identified users'' or SIUs.
Through this heuristic, we extracted $140$ users for ovarian cancer, and $41$ users for cervical cancer.
We assigned the label ``1'' to this subset of $\population$, while the rest of the users were labeled as ``0''.

We define two features matrix $X$ and $Z$ using the queries in $\queries$.
For each user, $X$ contains features extracted from queries issued after the first query mentioning the disease.
For example, the $i$-th row of $X^{ov}$ contains features extracted from all queries submitted by the $i$-th user in $\population^{ov}$ after searching for \textit{``ovarian cancer''} for the first time.
Conversely, $Z$ contains features extracted from all queries issued before the first query mentioning the disease.
A full list of features used in $X$ and $Z$ is reported in Table \ref{table:features-applications}.
The features matrix $Z$ is used by the classifiers introduced in Sections \ref{sub:applications:pre-screening} and \ref{sub:applications:predicting-statistics}.
Matrix $Z$ is comprised of queries mentioning symptoms and most common tokens, excluding stopwords, numbers, or names of the top one hundred websites in the US as ranked by Alexa (\url{http://alexa.com}). The latter was used so as to remove navigational queries. 
The number of tokens in $Z$ exceeded, for both datasets, fifty thousand.
In an effort to remove noise, we decided to keep only the top $q$ tokens;
$q$ was set to $2,000$ after empirical evaluation.
The feature matrix $X$ is used by the \perceptron to infer, for each user, their likelihood of being affected by cancer;
therefore, we also consider names of diseases, drugs, and US hospital as features.
Upon completion of the feature extraction phase, matrices $X^{ov}$, $X^{cr}$, $Z^{ov}$, and $Z^{cr}$ contain $7605$, $8766$, $2176$, and $2170$ features respectively.

As in Section \ref{sub:evaluation-plan:data-description}, we represent the state each user in $\population$ belongs to through the property matrix $P$.
The population statistic vector $\PropGlobal$ was obtained from the CDC \cite{us2015united}.
Similarly to Section \ref{sub:evaluation-plan:data-description}, the vector was normalized by the number of active Bing users in each state during the data period.

% \begin{figure*}[th]
%     \centering
%     \includegraphics[width=0.9\textwidth]{ovarian-cancer.pdf}
%     \caption{
%         pROC curves for ovarian cancer.
%         Each set of curves is associated with a value of the learning percentile $\delta$.
%         The performance of \classifiersugg under multiple values of $\theta$ are reported alongside the performance of the optimal and baseline classifiers.
%         Classifier $\theta = 0.95, \delta=0.90$ performs best (AUC = $0.776$, equivalent to $80.75\%$ of the optimal AUC).
%         Classifiers $\theta = 0.9$ and $\theta = 0.8$ lead to significant different classification outcomes (Wilcoxon signed-rank test, $p > 0.05$) for different values of $\delta$.}
%     \label{fig:proc-ovarian-cancer}
% \end{figure*}
% \begin{figure*}[th]
%     \centering
%     \includegraphics[width=0.9\textwidth]{cervix-cancer.pdf}
%     \caption{
%         pROC curves for cervical cancer.
%         The values of the learning percentile $\delta$ are reported above each set of curves.
%         The performance of \classifiersugg under multiple values of $\theta$ are shown alongside the optimal and baseline classifiers.
%         The classifier $\theta = 0.95, \delta=0.90$ achieves an AUC of $0.814$, which is equivalent to $84.62\%$ of the optimal AUC.
%         Classifier $\theta = 0.8$ lead to significant different classification outcomes (Wilcoxon signed-rank test, $p = 0.05$) for different values of $\delta$.}
%     \label{fig:proc-cervix-cancer}
% \end{figure*}

\subsection{Suggesting Medical Pre-Screening to Users}
\label{sub:applications:pre-screening}

Here, we introduce a classifier designed to identify search engine users who show signs of being potentially affected by cancer.
The classifier is designed to assess, for each searcher, their risk of developing cancer based on their query logs. The classifier is based on the labels inferred using the proposed algorithm, and uses past queries to assess if users will later be classified as suffering from the cancer of interest.

A logistic classifier \classifiersugg is trained to achieve the desired goal.
The classifier uses the feature matrix $Z$;
to obtain labels to train the system on, we proceed as follows:
first, we run the \perceptron on input $(X, P, \labels, \PropGlobal, \eta, \delta)$, where $X, P,$ and $\labels$ are as defined in Section \ref{sub:applications:data_description}, $\eta$ is set to $10,000$ for ovarian cancer and to $30,000$ for cervical cancer, and $\delta$ is varied between $0.95$ and $0.80$.
Then, once obtained the confidence vector $\likelihood$ for elements in $\population$, we extract users whose risk factor is in the \tth percentile of \likelihood, as well as those users whose risk factor is in the \lth percentile of \likelihood.
The former are used as positive training examples, while the latter are used as negative training examples.

Since we expect the number of users with no cancer to be greater than the number of users with cancer, we fix $\lambda = 3 (1 - \theta)$.
Therefore, the training set contains three negative examples for each positive example, somewhat mitigating the class imbalance probelm.
The weighting of each class was adjusted accordingly when training the classifier.

As a baseline, we consider a linear SVM trained solely on self-identified users.
This baseline was adapted from the classifier introduced by Yom Tov, et al. \cite{yomtov2015} to identify search engine users who have specific medical issues.
Specifically, we use SIUs as positive training examples and a sample of users from the reminder of the population as negative examples. Similarly to \classifiersugg, we sample three times the number of SIUs as negative examples.

\subsubsection{Results} % (fold)
\label{subsub:applications:pre-screening:results}

\begin{figure}[t]
    \centering
    \begin{subfigure}[t]{0.23\textwidth}
        \includegraphics[width=\textwidth]{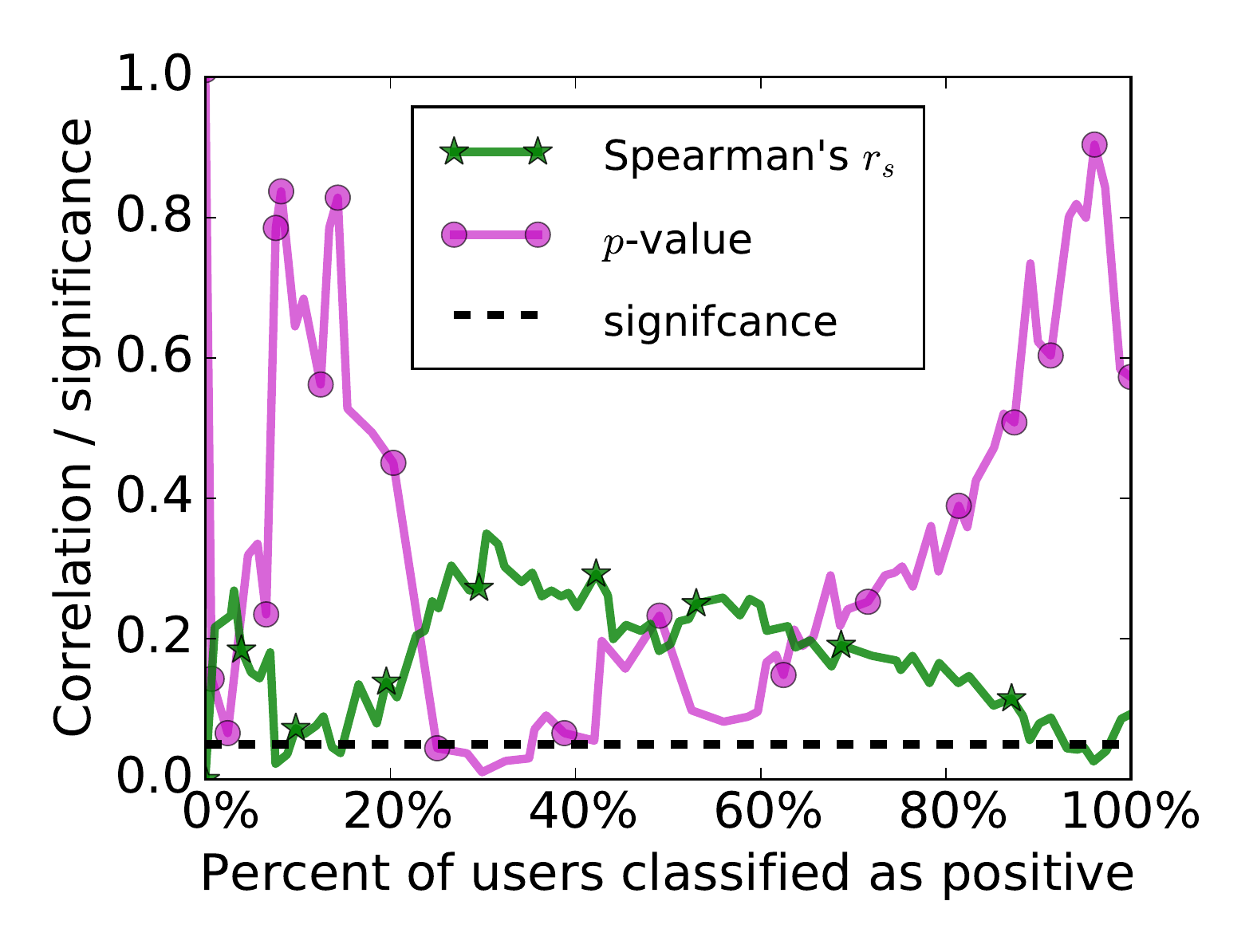}
        \caption{Ovarian cancer users}
        \label{subfig:regional_ovarian}
    \end{subfigure}
    ~ %add desired spacing between images, e. g. ~, \quad, \qquad, \hfill etc.
    \begin{subfigure}[t]{0.23\textwidth}
        \includegraphics[width=\textwidth]{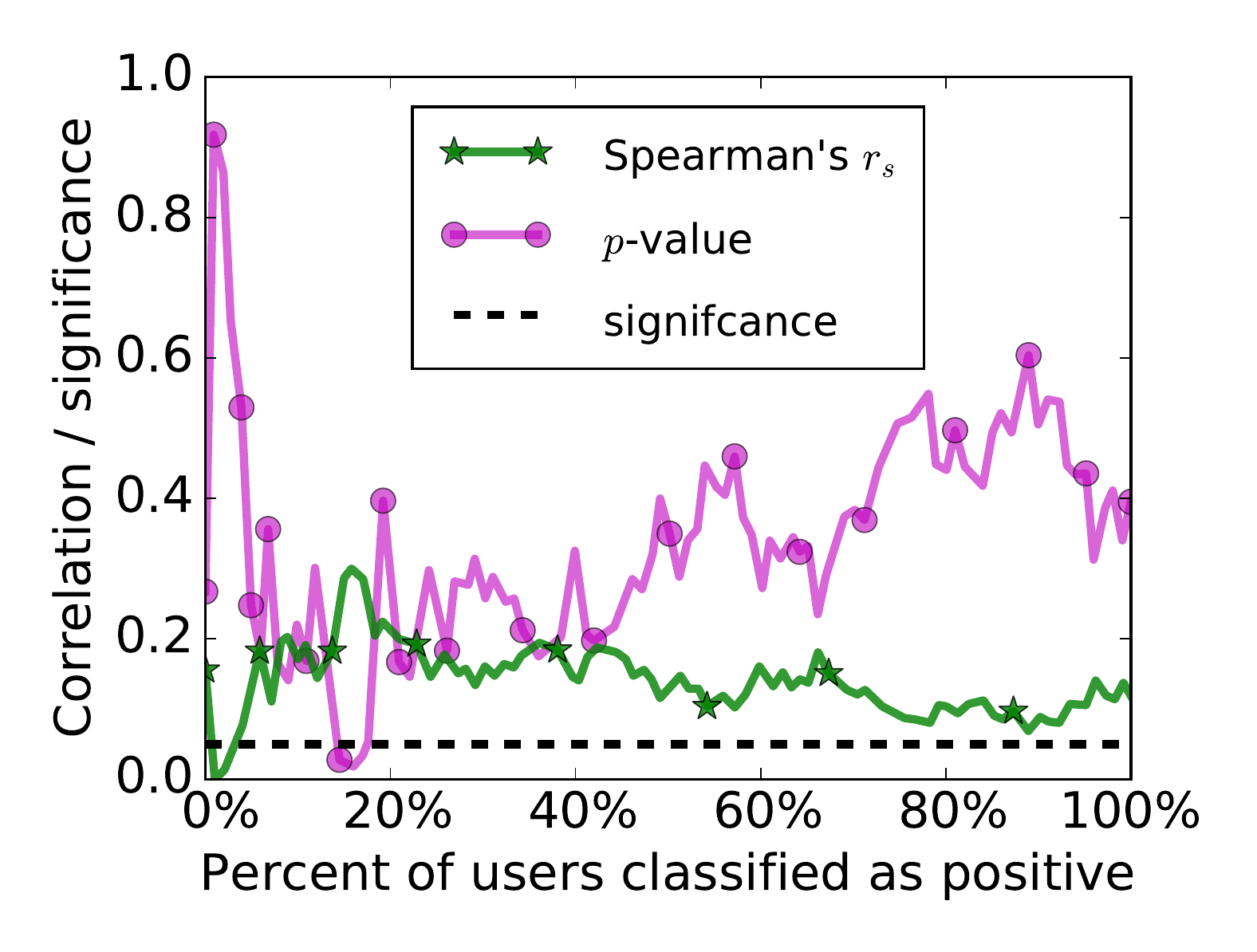}
        \caption{Cervical cancer users}
        \label{subfig:regional_cervical}
    \end{subfigure}
    \caption{In \textbf{\color{ForestGreen} green}, Spearman's rank correlation coefficient $\rho_s$ between the users identified by classifier \classifierreg and disease incidence as a function of users identified by \classifierreg.
    $p$-values are shown in \textbf{\color{violet} purple}.
    For ovarian cancer, \classifierreg achieve a statistically significant correlation ($\rho_s = 0.35$, $p < 0.05$) when $963$ users in $\population_{ov}$ are classified as positive
   (for cervical cancer: $\rho_s = 0.35$, $1,483$ individuals).
%     For cervical cancer, statistically significant correlation ($\rho_s = 0.30$) is obtained at $1,483$ individuals.
    %In both cases, $\delta = 0.90$, $\theta = 0.95$.
    }
    \label{fig:regional}
    \vspace{-14pt}
\end{figure}

Standard ROC methodology plots the fraction of correctly classified positive instances as a function of the fraction of incorrectly classified negative instances;
however, in this setting, such technique cannot be applied, as the true labels of the examples are not known. Instead, only a probability of the labels' correctness is known.

Techniques to adapt ROC analysis to probabilistic labels have been proposed in the literature;
in this work, we take advantage of the methodology introduced by Burl, et al. in \cite{burl1994automated}.
Let \probout be the probabilistic output of classifier \classifiersugg, $\probout = \{c_1, \ldots, c_n\}$.
Recall that $\likelihood = \{l_i\}_{i=1}^{n}$ (where $l_i \in [0, 1]$ for all $i$) is the likelihood of each element in the population of being in the cohort of interest, i.e., for this application, of suffering from cancer.
Then, for each decision threshold $\tau_i$ of classifier \classifiersugg, the following two quantities can be defined:
\begin{gather}
% \label{eq:pTPR}
    \text{pTPR}(\tau_i) = {(\Sigma_{j=1}^{i}l_j)}/{(\Sigma_{j=1}^{n}l_j)} \\
% \end{equation}
% \begin{equation}
% \label{eq:pFPR}
    \text{pFPR}(\tau_i) = {(\Sigma_{j=1}^{i}(1 - l_j))}/{(\Sigma_{j=1}^{n}(1 - l_j))}
\end{gather}
% \vspace{2pt}
The set of points $(\text{pTPR}(\tau_i), \text{pFPR}(\tau_i))$ for all values of $\tau_i$ define a curve in the ROC plane, which we refer to as probabilistic Receiving Operating Curve, or pROC.

A few observations can be made regarding any pROC curve.
First, we point out that, unlike in standard ROC analysis, the maximum AUC achievable by any \classifiersugg is less than one.
This is due to the fact that, even in case of perfect classification, the true and false positive rate are bounded by the probabilistic labels in \likelihood.
A corollary is that for any value of the false positive rate, the probabilistic true positive rate of the classifier is a lower bound on the actual true positive rate.
This explains why, in Figures~\ref{fig:proc}, the optimal pROC curve---which is obtained when the labels are known with complete accuracy---is described by a curve rather than the segments $[(0,0), (0, 1)]$ and $[(0,1), (1, 1)]$.

Results of classifier \classifiersugg on the dataset of ovarian cancer and cervical cancer users are reported in Figures \ref{fig:proc}a and \ref{fig:proc}b.
Each run is evaluated using five-fold stratified cross validation.
For both dataset, we present three groups of pROC curves, each one associated with a different value of the learning percentile $\delta$.
Each group consists of the optimal classification pROC curve, four curves associated with four values of training percentile $\theta$, and the pROC curve of the baseline classifier.

First, we note that all configuration of the classifier perform substantially better than the baseline.
This is to be expected, as the baseline classifier is trained on very few examples.
Furthermore, we observe that the baseline classifier for cervical cancer is decisively worse than the baseline classifier for ovarian cancer.
We believe that the phenomenon is due to the fact that the number of self-identified cervical cancer users is substantially smaller than the number of self-identified ovarian cancer users. Thus, both a small number of SIUs and the population-level data are needed to correctly identify users.

Third, we note that, for all values of $\delta$, not all classifiers are significantly different from each other (Wilcoxon signed-rank test, $p < 0.05$) with the exception of $\theta = 0.80$.
This is a desirable outcome: since the size of users in the cohort of interest is unknown, a classifier that is resistant to small variations of the tuning parameters is beneficial.

%3. for lower value of theta, runs are significantly different for different values of delta. This indicates that they are less robust due to noise in the training examples.
Lastly, we study the differences in classification outcomes for fixed values of $\theta$.
We notice that, once again, there are no significant differences between $\theta = 0.99$ and $\theta = 0.95$.
For the cervical cancer dataset, this is the case for $\theta = 0.90$ as well.
However, we observe that, as $\theta$ decreases, the performance of the classifier becomes less stable.
In particular, the classification outcomes associated with different values of $\delta$ significantly differ (Wilcoxon signed-rank test, $p > 0.05$).
This event is likely to be caused by that fact that, as $\delta$ and $\gamma$ decrease, users who are not affected by cancer might be part of the learning or training percentile, which naturally decreases the accuracy of the classifier.

\subsection{Predicting Disease Distribution}
\label{sub:applications:predicting-statistics}

In this section we show how the probabilistic labels computed by the proposed algorithm can be exploited to predict the incidence of diseases in areas for which it is not known.
Specifically, we introduce a logistic classifier \classifierreg that identifies search engine users affected by the disease of interest in states with unknown incidence.
The incidence in each region can then be determined by dividing the number of users identified by the number of active search engine users.
Thus, we can infer disease incidence in areas where it is unknown, which is an important utility for epidemiologists interested in the spread of a disease.

%We do not attempt to predict the incidence directly, as the labels returned by the \perceptron are at an individual level.

The procedure to train \classifierreg is not dissimilar from the one used to train \classifiersugg (Section \ref{sub:applications:pre-screening}).
However, unlike \classifiersugg, matrices $X$ and $Z$ were combined to train the system.The learning percentile $\delta$ and the training percentile $\theta$ were set to $0.90$ and $0.95$, respectively; these value were chosen based on the results described in Section \ref{subsub:applications:pre-screening:results}. The classifier is evaluated using the dataset introduced in Section \ref{sub:applications:data_description} under five-fold cross validation.

\subsubsection{Results} % (fold)
\label{subsub:applications:predicting-statistics:results}

The results of the classifier \classifierreg on the ovarian and cervical cancer datasets are shown in Figure \ref{fig:regional}.
We report Spearman's rank correlation coefficient $\rho_s$ between the number of users identified by classifier \classifierreg and disease incidence as reported by the Center of Disease Control as a function of the percentage of users identified as positive by \classifierreg.
Before calculating the correlation, counts of users identified by the classifier in each state were normalized by the number of total search engine users in the state.

For both datasets, the classifier is able to obtain a statistically significant  correlation (Spearman's rank correlation test, $p < 0.05$) between the normalized number of users identified and the incidence of the disease.
\classifierreg reaches the highest correlation of $\rho_s = 0.35$ when $30\%$ of users are labeled as positive on the ovarian cancer dataset (Figure \ref{subfig:regional_ovarian});
similarly, it obtains a correlation of $\rho_s = 0.30$ when $16\%$ of users are label as positive on the ovarian cancer dataset (Figure \ref{subfig:regional_cervical}).
We note that the large difference in percentage of positively label users between the two datasets is mostly due to the fact that $\population^{ov}$ and $\population^{cr}$ are of different sizes;
in fact, \classifierreg classifies a similar number of users as positive at the at the point of maximum correlation: $963$ for ovarian cancer dataset and $1,483$ for the cervical cancer dataset.
The smaller difference in terms of individuals classified as positive is more consistent with the US incidence provided by the CDC, which is similar for the two diseases.

We also point out that, for both datasets, correlation follows a similar pattern:
when \classifierreg labels very few users (left side of Figures \ref{subfig:regional_ovarian} and \ref{subfig:regional_cervical}) the correlation with CDC data is low and not significant;
then, as the number of positively classified users increases the correlation value improves up to reaching statistical significance.
However, it declines and looses significance as the number of users classified as positive approaches the size of the population (right side of Figures \ref{subfig:regional_ovarian} and \ref{subfig:regional_cervical}).

Finally we note that, while the correlation values are modest, previous research \cite{yomtov2015} that used only SIUs found a correlation of $0.45$ between HIV incidence and number of users. Thus, our correlations are close to those achieved using only users who are known to be suffering from a condition.

\section{Additional observations}

\begin{table}[t]
%     \vspace{-12pt}
    %\small
    \centering
    {\def\arraystretch{1}\begin{tabularx}{0.94\columnwidth}{|l|c|c|X|}
        \hline
        & \textbf{SIUs} & \textbf{non-SIUs} & \textbf{SIUs to non-SIUs} \\
        \hline
        Cervical & 0.942 & 0.918 & \multicolumn{1}{c|}{0.934} \\
        Ovarian  & 0.938 & 0.926 & \multicolumn{1}{c|}{0.936} \\
        \hline
    \end{tabularx}
    \vspace{-8pt}
    }
    \caption{Average cosine similarities among SIUs, among non-SIUs identified as suffering from the condition of interest, and between SIUs and non-SIUs.}
    \label{table:similarities}
    \vspace{-12pt}
\end{table}

\subsection{Stability of the Algorithm} % (fold)
\label{sub:stability_of_the_algorithm}

As the \perceptron algorithm attempts at separating positive and negative users in \population with hyperplane \weights, it is natural to ask whether the solution it identifies for a given dataset is stable.
To answer this question, we ran Algorithm~\ref{alg:perceptron} ten times and measured (\textit{i}) the rank correlation between the scores of users from any two runs and (\textit{ii}) the inter-run agreement between all runs.
Results show that, for the dataset introduced in Section~\ref{sub:evaluation-plan:data-description}, the Spearman's rank correlation between any two runs is at least 0.8 (statistically significant, $p < 0.05$); furthermore, the inter-run agreement is 0.73, which suggests high agreement among runs.
Similar results are obtained for the datasets introduced in Section~\ref{sub:applications:data_description}.
This shows that the \perceptron achieves very similar prediction despite the sampling process in Algorithm~\ref{alg:perceptron}, lending additional credence to the hypothesis that users identified by the algorithm do indeed share the trait of interest.

\subsection{The Similarity of SIUs to Other Patients}
\label{sub:applications:similarity-self-identifed}

\begin{figure}[t]
    \centering
    \begin{subfigure}[t]{0.23\textwidth}
        \includegraphics[width=\textwidth]{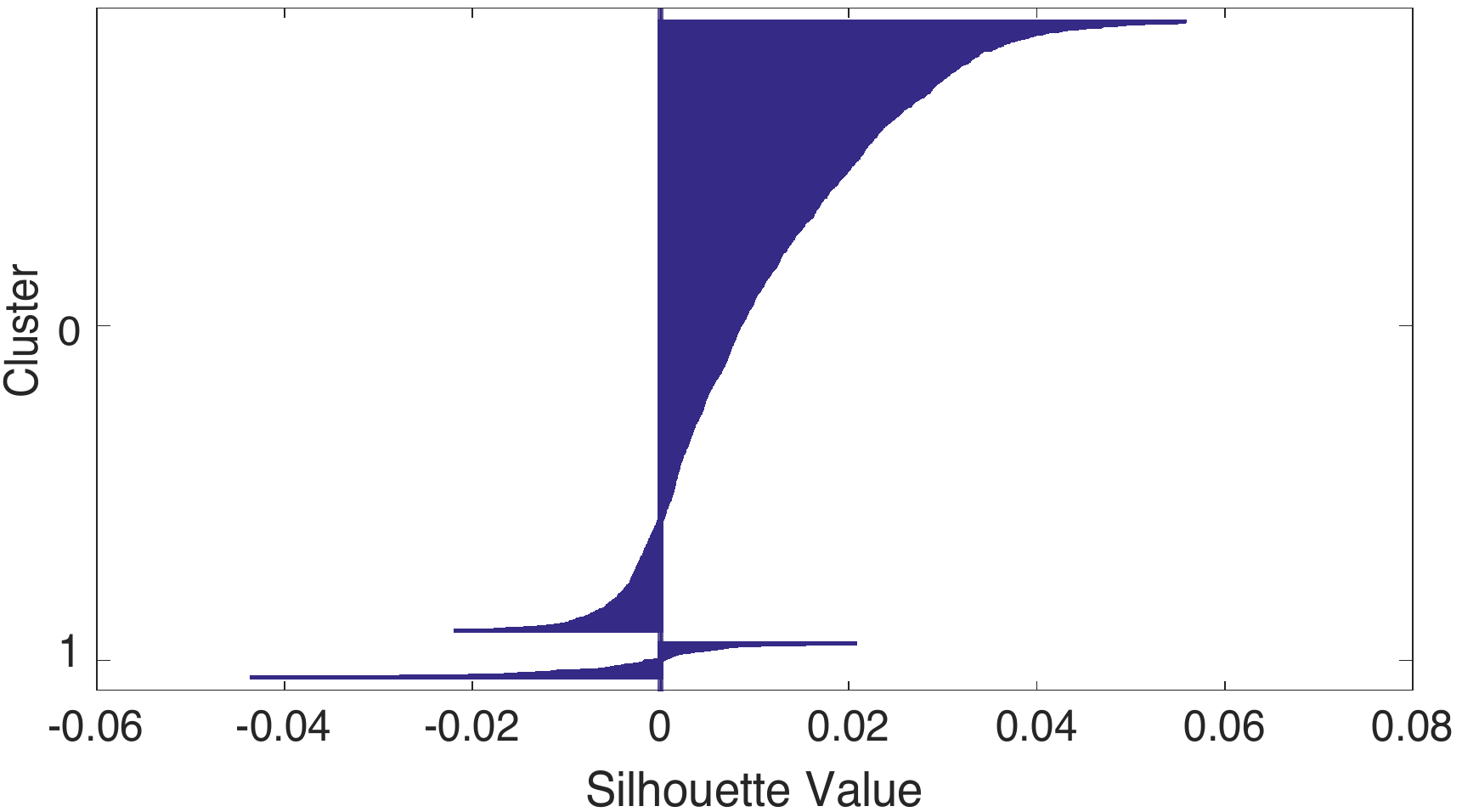}
        \caption{Ovarian cancer users}
        \label{fig:silhouette_ovarian}
    \end{subfigure}
    ~ %add desired spacing between images, e. g. ~, \quad, \qquad, \hfill etc.
    \begin{subfigure}[t]{0.23\textwidth}
        \includegraphics[width=\textwidth]{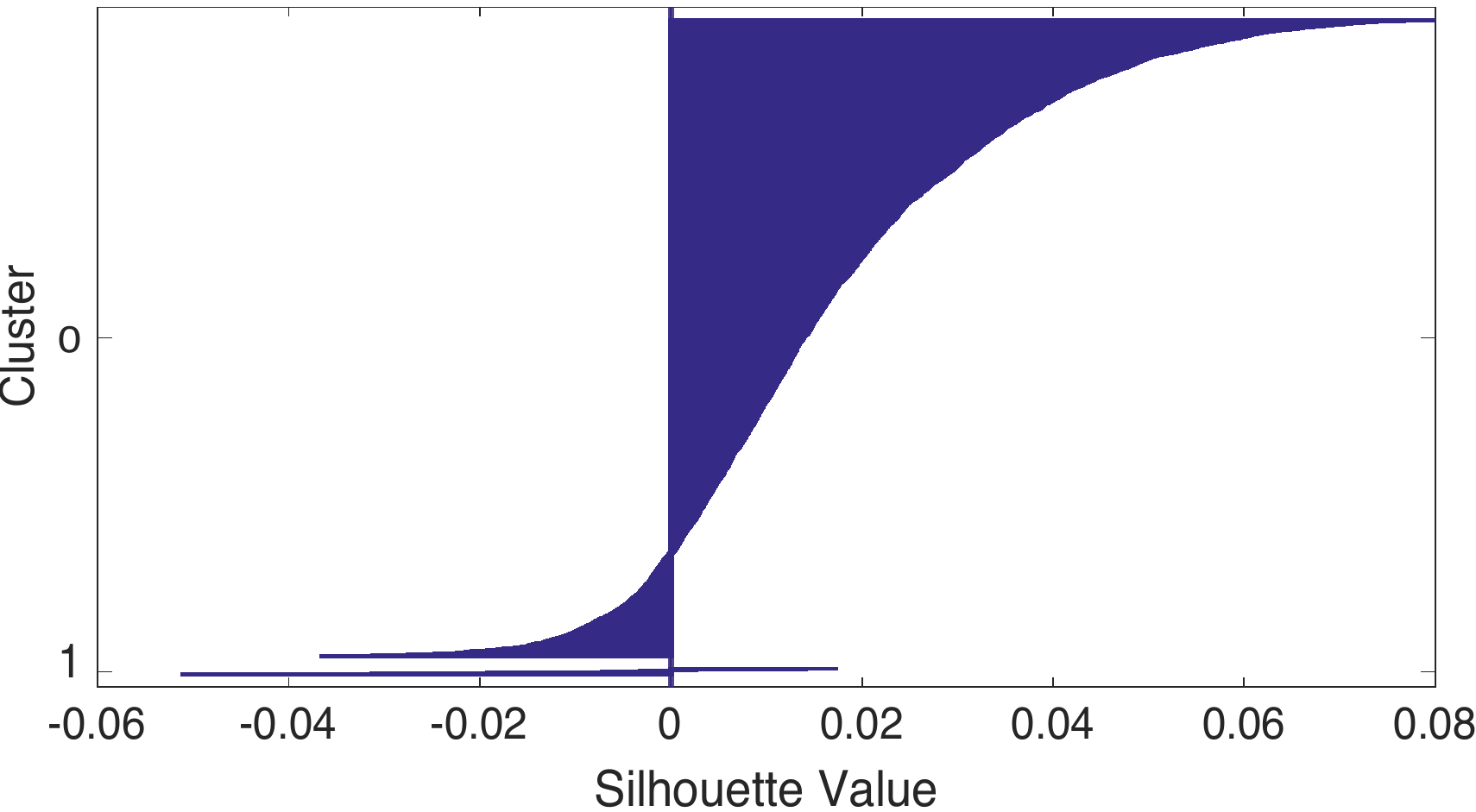}
        \caption{Cervical cancer users}
        \label{fig:silhouette_cervical.pdf}
    \end{subfigure}
    \caption{Silouette charts for the similarity between SIUs (denoted by ``1'') and non-SIUs identified as patients (denoted by ``0''). Negative values imply similarity between classes and positive values dissimilarity. }
    \label{figure:silhouette}
    \vspace{-12pt}
\end{figure}

Previous work \cite{paparrizos2016,yomtov2015} used anonymous self-identified users (SIUs) to identify behaviors associated with other users suffering from the condition of interest. Having noted the poor performance of predicting diseases using only SIUs, in this section we ask whether this performance could be because the behavior of SIUs is not representative of other users suffering from the condition of interest.

We compared all SIUs of a condition (ovarian or cervical cancer) with non-SIUs in the top 10\% of users found by the Perceptron run at a 95\% threshold (i.e., $\delta = 0.95$).

Table \ref{table:similarities} shows the average cosine similarity (computed from $X$) between users within the two classes and between users of different classes. As the table demonstrates, SIUs are most similar among themselves. Non-SIUs are least similar, and the similarity between non-SIUs and SIUs is in between the two user classes. A different way to observe these similarities is through the silhouette graphs \cite{kaufman2009} shown in Figure \ref{figure:silhouette}. As the graphs show, there are some similarities between groups (SIUs vs. non-SIUs), as demonstrated by the negative values on the charts, but also significant amounts of dissimilarities (positive values on the graphs).

These results imply that SIUs are different from other users identified as suffering from the conditions of interest. This lends additional support to the importance of using the proposed algorithm to identify additional users beyond the small number of self-identified users.

%!TEX root = resolving-missing-labels.tex

\section{Conclusion}
\label{conclusions}

In this paper, we introduced a novel algorithm for identifying cohorts of interest among internet users.
Our approach exploits a small set of users whose membership to the cohort of interest is known (e.g., they self identified themselves) alongside statistics on the entire population.
The algorithm was validated on a political dataset of tweets in Section \ref{sec:evaluation}.
Then, in Section \ref{sec:applications}, we introduced two applications of the proposed algorithm.
First, we discussed a classifier designed to pre-screen for specific forms of cancer using search engine queries.
This system could be of great help in detecting  diseases that have a set of nonspecific symptoms, no screening test, or may have increased risk if not diagnosed early.
% However, further research is required to validate whether the sensitivity and specificity of this approach is high enough to be of practical purpose.
The second application we investigated dealt with predicting the incidence of a disease in regions in which it is not known.
The proposed classifier would be of high value in cases where the incidence of a disease is too low to be measured in a specific region by traditional surveillance methods, or when a disease is spreading within a population.
Alternatively, such system could also be helpful in those cases where, for technical reasons, incidence of a disease was not reported.

An important observation stemming from our work is that, when studying anonymous users, SIUs are insufficiently representative  of the population.
This is both because of the dearth of SIUs, but also, possibly, because there is something unique in the behavior of those users who self-identify.
However, SIUs are crucial for identifying the cohort. This observation means that algorithms such as the one proposed herein are needed for the study of anonymous users.

% Future directions of our research include directly validating our pre-screening algorithm with a clinically validated cohort. Other directions include the exploration of the use of the proposed framework in other domains (e.g., marketing).
%Furthermore, we plan to improve the classifier introduced in Section \ref{sub:applications:pre-screening} to take advantage of unique characteristics of query logs.
%Finally, characterizing the difference between self-identified users and individuals extracted by the \perceptron will be also considered in future studies.

\section{Acknowledgments}

We are grateful to Yishay Mansour for his insights and enlightening conversations throughout this work.

%\clearpage

\begin{normalsize}

\end{normalsize}

%----------------------------------------------------------------------------------------

\end{document}